\documentclass[10pt, conference]{IEEEtran}
\ifCLASSINFOpdf
\else
\fi
\pdfoutput=1
\usepackage{algorithm}
\usepackage{algorithm,algpseudocode}

\usepackage{amsmath}    
\usepackage{amssymb}
\usepackage{graphicx}   
\usepackage{array,multirow}  
\usepackage[scientific-notation=true]{siunitx}
\usepackage{verbatim}   
\usepackage{color}      
\usepackage{subfig}  
\usepackage[draft]{hyperref}   
\usepackage{mathtools}
\usepackage[normalem]{ulem} 
\newcommand{\stkout}[1]{\ifmmode\text{\sout{\ensuremath{#1}}}\else\sout{#1}\fi}
\usepackage{soul} 
\usepackage{kantlipsum}
\usepackage{comment}
\usepackage{tabularx}
\allowdisplaybreaks

\def \Tfic {\textsc{T}_{f}^{\SFCidx i}}
\def \tfic1 {\textsc{T}^{ci}_{f: f_i \notin \{f^c_s,f^c_d\}}}

\def \ncore {n^{\textsc{core}}}

\def \ncoref {n^{\textsc{core}}_{f}}

\def \NFV         {\textsc{nfv}}
\def \NFVI {\textsc{nfv}}

\def \ServiceChainSet {C}
\def \SFCidx {c}
\def \SFCset {C}
\def \ServiceChainIdx {c}

\newcolumntype{P}{>{\raggedright\arraybackslash}p{3cm}}
\newcolumntype{M}{>{\raggedright\arraybackslash}m{3cm}}

\begin{document}

%
\title{Virtual-Mobile-Core Placement for Metro Network
\large{\color{green}To appear in proceedings of IEEE NetSoft 2018}}

\author{\IEEEauthorblockN{Abhishek Gupta\IEEEauthorrefmark{1},
Massimo Tornatore\IEEEauthorrefmark{1}\IEEEauthorrefmark{3}, Brigitte Jaumard\IEEEauthorrefmark{2}, and
 Biswanath Mukherjee\IEEEauthorrefmark{1}}
\IEEEauthorblockA{\IEEEauthorrefmark{1}University of California, Davis, USA \ \ 
\IEEEauthorrefmark{2}Concordia University, Canada \ \
\IEEEauthorrefmark{3}Politecnico di Milano, Italy \\
Email: \IEEEauthorrefmark{1}\{abgupta,mtornatore,bmukherjee\}@ucdavis.edu  \IEEEauthorrefmark{2}bjaumard@cse.concordia.ca  \IEEEauthorrefmark{3}massimo.tornatore@polimi.it} }


%


\maketitle

\begin{abstract}
Traditional highly-centralized mobile core networks (e.g., Evolved Packet Core (EPC)) need to be constantly upgraded both in their network functions and backhaul links, to meet increasing traffic demands. Network Function Virtualization (NFV) is being investigated as a potential cost-effective solution for this upgrade. A virtual mobile core (here, virtual EPC, vEPC) provides  deployment flexibility and scalability while reducing costs, network-resource consumption and application delay. Moreover, a distributed deployment of vEPC is essential for emerging paradigms like Multi-Access Edge Computing (MEC). In this work, we show that significant reduction in network-resource consumption can be achieved as a result of optimal placement of vEPC functions in metro area. Further, we show that not all vEPC functions need to be distributed. In our study, for the first time, we account for vEPC interactions in both data and control planes (Non-Access Stratum (NAS) signaling procedure Service Chains (SCs) with application latency requirements) using a detailed mathematical model. 
\end{abstract}

%
\IEEEpeerreviewmaketitle

\section{Introduction}
\label{intro}
Mobile network operators have to manage increasing traffic from bandwidth-intensive applications. This creates challenges for the Evolved Packet Core (EPC) (i.e., the current state-of-the-art mobile core network) where complex dependencies among various EPC elements lead to scalability issues as traffic demand increases and results in frequent upgrades of costly proprietary hardware.  

Network Function Virtualization (NFV) virtualizes network functions (called Virtual Network Functions, VNFs) and runs them on commodity hardware. Also, EPC functions can be virtualized into VNFs to provide cost savings, flexibility, and ease of deployment. We refer to EPC implemented using VNFs as virtual EPC (vEPC). In order to understand possible VNF inter-dependencies and interactions in vEPC, we must discuss what a traditional EPC is and how it functions.

\begin{figure}
\center
 \includegraphics[width=.45\textwidth, scale=1]{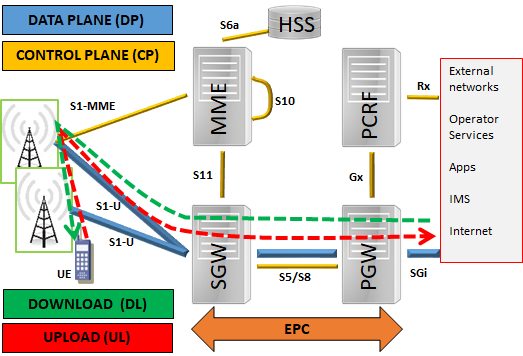}
 \caption{Evolved Packet Core (EPC).}
  \label{fig:epc}
\end{figure}

Evolved Packet Core (EPC) is an end-to-end IP-based mobile core network infrastructure. Fig. \ref{fig:epc} shows a high-level functional view of an EPC, its functional entities and interfaces. Delineation between control and data planes is an important aspect of EPC and will be discussed further. Mobility Management Element (MME), Policy and Charging Rules Function (PCRF), and Home Subscriber Server (HSS) are EPC control plane elements. Serving Gateway (SGW) and Packet Data Network Gateway (PGW) are EPC data plane elements but also have tightly coupled control plane functions. In this work, we consider vEPC to have VNFs for MME, PCRF, HSS, SGW and PGW. 

User Equipment (UE) connects to the Internet (or in general to an external network) through EPC. The data path (bearer) to the Internet for a UE is setup as a result of control plane signaling procedures called Non-Access Stratum (NAS) procedures. NAS is a layer-3 protocol for mobile networks, responsible for management and modification of bearers. The types of EPC functions involved in a NAS procedure depend on the UE signaling event. Thus, there are different NAS procedures, such as Attach, Dedicated bearer setup, X2-based handover, S1-based handover etc. as detailed in \cite{nas_procs}. NAS procedures form a critical part of control plane signaling. A slight increase (e.g.,1\%) in the control plane signaling can result in a significant decrease (70\%) in data plane capacity \cite{rajan_nas_paper}. This happens because SGW while handling all data plane traffic is also involved in 33\% \cite{rajan_nas_paper} of all control plane transactions. A highly-loaded SGW can become a bottleneck which affects data path throughput and  increases control plane latency.

\begin{figure*}
  \centering
  \begin{tabular}{llrr}
  \multicolumn{2}{c}{\multirow{2}{*}[1.5cm]{\subfloat[][Simplified NAS Attach Procedure\cite{nas_procs}]{\label{fig:a}\includegraphics[width=.4\textwidth, height=4cm, scale=1]{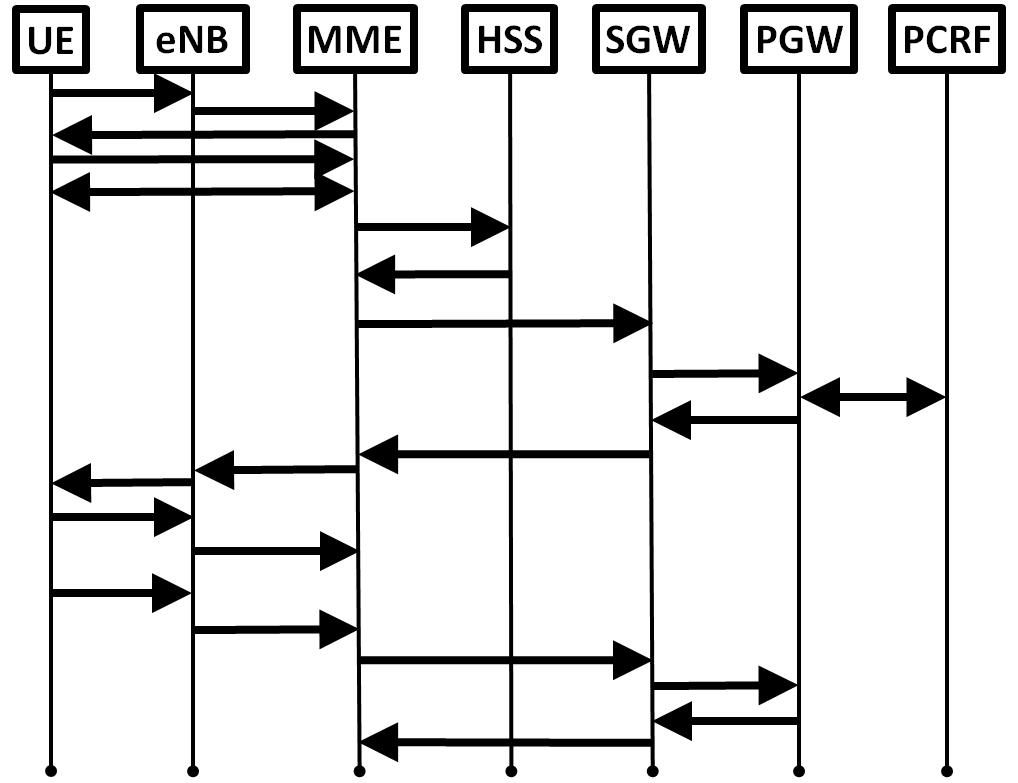}}}}
    &  \multicolumn{2}{c}{\subfloat[][Attach Control Service Chain (CSC)]{\label{fig:b}\includegraphics[width=.5\textwidth, scale=1]{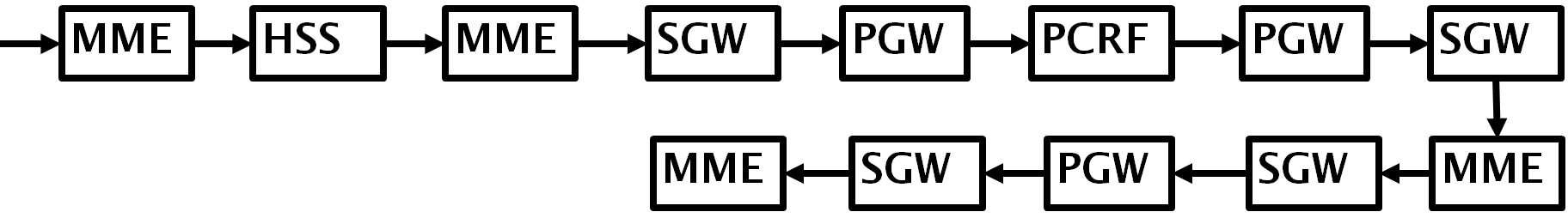}}} \\ 
   \multicolumn{2}{c}{} & \subfloat[][Data Service Chains (DSCs)]{\label{fig:c}\includegraphics[width=.20\textwidth, scale=1]{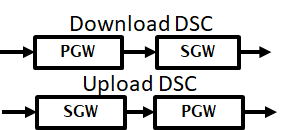}}
   &  \subfloat[][Stateful VNFs in Attach CSC]{\label{fig:d}\includegraphics[width=.25\textwidth, scale=1]{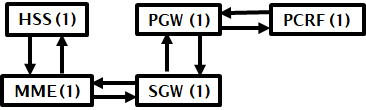}} \\ 
  \end{tabular} 
  \caption{NAS Procedure, control and data Service Chains (SCs)}
  \label{ctrl_data_details}
\end{figure*} 

Figure \ref{ctrl_data_details}\subref{fig:a} shows the NAS Attach Procedure, where all EPC functions are involved in control signaling. Here, signaling proceeds to next function only after it has been processed by current function. So, control plane signaling can be characterized as a chained sequence of interactions between EPC functions. We refer to this chained sequence of interactions in the control plane as a Control Service Chain (CSC)\footnote{The term ``Service Chain''(SC)\cite{ietf_sc} is usually used for value-added services (Firewall, Video Optimization etc.). We use ``Service Chain'' here since an ordered sequence of functions is similar to chained sequence of interactions between functions.}. Fig. \ref{ctrl_data_details}\subref{fig:b} shows the corresponding CSC for attach procedure. 

NAS attach procedure completion results in a data path (bearer) for UE to Internet. Note that control-plane latency requirements for each NAS procedure depends on whether a default or dedicated bearer is setup. The UE upload traffic reaches eNodeB and is directed to SGW. SGW is the mobility anchor for eNodeB's, and hence, upload traffic has to traverse SGW first, then PGW to reach Internet as shown in Fig. \ref{fig:epc}. If a UE downloads data, SGW and PGW are traversed in reverse order. Fig. \ref{ctrl_data_details}\subref{fig:c} shows the Data Service Chains (DSCs) for upload and download.

An important aspect of EPC functions is that they are stateful. They need to maintain session information, i.e., maintain information regarding UE and bearer state and internal database states, which might be different between instances of the same function. For example, there could be two VNF instances for MME, where the first MME (MME(1)) holds session information for a different set of UE's from the second MME (MME(2)). For NAS Attach procedure shown in Fig. \ref{ctrl_data_details}\subref{fig:a} the EPC function interactions happen with the same instance of EPC functions as seen in Fig. \ref{ctrl_data_details}\subref{fig:d} (1 indicating a single instance). The required number of instances of a function will depend on the NAS procedure. This is an important distinction between EPC SCs and traditional (e.g., value-added) SCs where VNFs can be stateless. In this work, we solve vEPC placement with stateful VNFs unlike previous works (e.g., our work in \cite{gupta_gc17}\cite{gupta_jsac18}) for traditional SCs. We account for stateful VNFs by tracking the VNF instance in our model. 

In this paper, we propose an Integer Linear Program (ILP) to reduce bandwidth consumption in metro core while accounting for control and data plane interactions of UEs and satisfying application latency requirements through the optimal placement of vEPC function replicas. This model allows us to show that having distributed vEPC replicas reduces bandwidth consumption and not all vEPC functions need to be distributed.  

The rest of this study is organized as follows. Section \ref{netw_arch} describes our network architecture. Section \ref{rel_work} summarizes existing literature on vEPC placement problem and remarks on the novelty of this study. Section \ref{prob_desc} provides details on problem definition, input parameters and ILP model. Section \ref{num_examples} and \ref{concl} details illustrative examples and conclusion.

\section{Network Architecture}
\label{netw_arch}
We assume a metro-core mesh topology connected to 2 aggregation rings to be our network architecture as shown in Fig. \ref{fig:results}\subref{fig:a}. Traffic from UEs is aggregated at Traffic Aggregation Points (TAPs). Aggregated UE traffic forms traffic flows that originate at a TAP and terminate at an application gateway or vice-versa. Application gateways in our architecture are peering points of the mobile core with other networks. For example, video traffic will go to application gateway that peers with the network that has the video content.

If all applications require traffic to be routed to mobile core to reach desired application gateway, network-resource consumption will increase with traffic demand. To avoid routing to mobile core, we need to host applications closest to the edge, i.e., in aggregation rings. So, in Fig. \ref{fig:results}\subref{fig:a} we include two Multi-Access Edge Computing (MEC)\footnote{MEC is an emerging paradigm which aims to reduce network-resource consumption and improve application latency by deploying cloud-based IT services at network edge.} \cite{etsi_mec} nodes for hosting applications.  




\begin{figure*}
  \centering  
  \includegraphics[width=0.75\textwidth, scale=1]{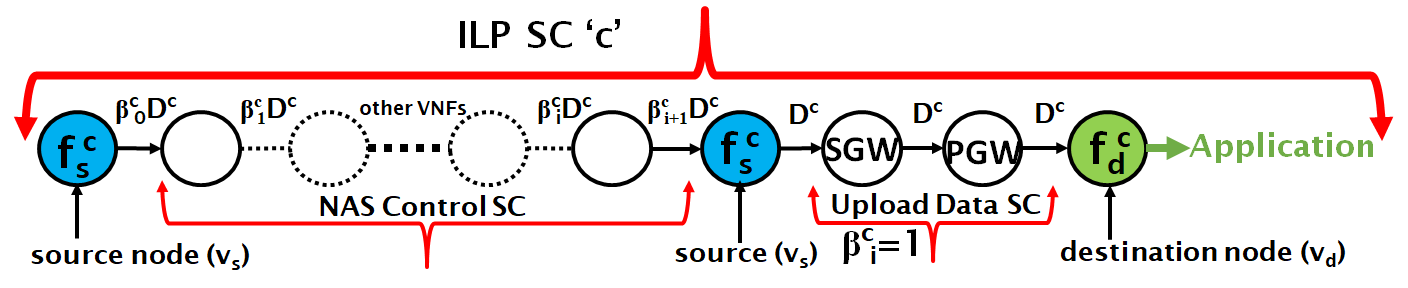}  
  \caption{Application flow requesting upload with NAS procedure}
  \label{fig:modeling_details}
\end{figure*}

\section{Related Works}
\label{rel_work}
Only few studies already exist for the vEPC-placement problem. Ref. \cite{baumgartner_vepc} was the first to address the vEPC placement problem. Author's model vEPC functions as service chains but do not account for NAS procedures in control plane and upload and download in data plane, modeling service chaining explicitly, stateful nature of vEPC elements and application latency. Ref. \cite{baras_vepc} proposes combined network function and vEPC placement while accounting for vEPC function mapping to eNodeB. A mathematical model and heuristic are proposed, however this work also has same set of limitations as Ref. \cite{baumgartner_vepc}. 

Other works, while not solving vEPC placement directly, provide useful insights. Ref. \cite{rajan_nas_paper} emulates vEPC functions and shows that SGW is a bottleneck in EPC, and proposes a better functional design for vEPC. We utilize this insight in our model as stated in Section \ref{intro}. Ref. \cite{topology_paper} demonstrates virtual SGW and PGW placement using a simulation framework, but does not provide details of the VNF-placement algorithm. Ref. \cite{taleb_relocation} focuses on minimizing SGW relocations without considering control and data plane interactions of other EPC network functions. 


To the best of our knowledge, this is the first attempt to account for control and data plane interactions and application latency for optimal placement of vEPC replicas in a metro network.

\section{Problem Description}
\label{prob_desc}
A mobile network operator requires an EPC for connecting user equipment (UE) to Internet. To provide services, EPC network functions have to engage in control signaling (chained requests or Control SC (CSC)) for data path (Data SC (DSC)) setup. We develop a Integer Linear Programming (ILP) model for optimal placement of vEPC functions which accounts for latency requirements of applications, control and data plane interactions and statefulness of EPC VNFs, number of VNF replicas available and nodes allowed to host VNFs.

\subsection{Problem Statement}
\label{prob_stat}
Given a network topology, capacity of links, a set of network nodes with NFV support (NFV nodes), compute resources at NFV nodes, number of NFV nodes used, aggregated traffic flows using a NAS procedure and requesting an application, and latency requirements for application and control signaling, we determine the placement of vEPC VNFs and traffic routing to minimize network-resource (bandwidth) consumption.  

\subsection{Input Parameters}
\label{input_param}

\noindent
\begin{tabular}{m{.7cm} m{7.5cm}}
$G$ &  Physical topology of backbone network \\
& $G= (V, L)$ with $V$: node set  and $L$: link set  \\
$\mathcal{SD}$ & Set of source-destination $(v_s,v_d)$ pairs \\
$V^{\NFV}$ &  $\subseteq V$ Set of nodes that can host VNFs (NFV nodes) \\
$\ncore$ & Number of CPU cores present in a NFV node \\
$F$ & Set of VNFs, indexed by $f$ \\
$R_{f}$ & Maximum number of replicas of VNF $f$ \\
$\ncoref$ & Number of CPU cores per Gbps for function $f$ \\
$\ServiceChainSet$ & Set of chains, indexed by $c$ \\
$n_c$ &  Number of VNFs in SC $\ServiceChainIdx$ \\
$\chi_f^c$ & Number of instances of VNF $f$ required by $c$ \\
$\mathcal{SD}_c$ & Source-destination $(v_s,v_d)$ pair for SC $c$ \\
$f^c_s$ & VNF for $v_s$ in SC $c$\\
$f^c_d$ & VNF for $v_d$ in SC $c$\\
$D^c$ & Traffic demand for SC $\ServiceChainIdx$ \\
$\sigma_i(\ServiceChainIdx)$ & ID of $i$th VNF in SC $\ServiceChainIdx$, \\
&   \hspace*{1.cm} where $f_{\sigma_i(\ServiceChainIdx)} \in F$ \\
$\Tfic$ & VNF ID ($f$) of the $i$th VNF in SC $c$
\end{tabular}

\begin{tabular}{m{.7cm} m{7.5cm}}
$\theta_i(\ServiceChainIdx)$ & Instance used for $i$th VNF with ID $f$ in SC $\ServiceChainIdx$, \\
&   \hspace*{1.cm} where $\theta_i(\ServiceChainIdx) \in \{1, 2, \dots, \chi_f^c\}$ \\
$\beta_{i}^{c}$ & Fraction of $D^c$ between $i$ and $(i+1)$th VNF \\
& \hspace*{1.cm} for SC $\ServiceChainIdx$ \\
${\Delta}_{\ell}^{\textsc{prog}}$ & Propagation latency for link $\ell$ \\
${\Delta}_{f}^{\textsc{proc}}$ & Processing latency for function $f$ \\
$L_c$                                    & Latency requirement for SC $c$
\end{tabular}


\subsection{Variables}
\label{ilp_var}

\noindent
\begin{tabular}{m{.3cm} m{8.1cm}}
$x_{v i}^{c j}$ & 1 if instance $j$ of $i$th function $f_i$ of c is located \\
& \hspace*{.5cm} in $v \in V^{\NFV}$ (responsible for stateful VNFs)\\
& 1 if $i$th function $f_i \in \{f_s^c,f_d^c\}$ is located in $v \in \mathcal{SD}_c$ \\
& 0 otherwise \\
$x_{vf}$ & 1 if function $f$ is located in $v \in V^{NFV}$; 0 otherwise \\
$y_{i\ell}^c$ & 1 if $\ell$ is on the path from location of $f_i$ \\
& to location of $f_{i+1}$, 0 otherwise.
\end{tabular}


\subsection{Problem Modeling}
\label{prob_form}

We consider aggregated traffic at Traffic Aggregation Points (TAPs) in $G$. For upload $v_s$ is a TAP and $v_d$ is an application gateway (opposite for download).

We consider that application flows request upload or download for an application and may or may not require a NAS procedure. To simplify formulation, we define a SC $c$ in our ILP as consisting of  $(v_s,v_d)$ pairs, NAS Control SC, Upload/Download Data SC and application as shown in Fig. \ref{fig:modeling_details}. Our ILP considers SC $c$ as a sequence of VNFs. This also includes $(v_s,v_d)$ which are source-destination nodes and are not VNFs. We create functions $f_s^c$ and $f_d^c$ to represent source $v_s$ and destination $v_d$ respectively. VNFs $f_s^c$ and $f_d^c$ have the only requirement of being located at $v_s$ and $v_d$ respectively.

Fig. \ref{fig:modeling_details} shows an application flow for upload with NAS procedure. Here, control traffic originates from VNF $f_s^c$ (which represents source node $v_s$) and traverses NAS CSC. After the traversal, control traffic reaches $f_s^c$ to notify that data path is setup. Since application traffic requests upload, the upload DSC is to be traversed on data path to destination (here, application gateway). Total latency requirement $L_c$ for SC $c$ is ``control plane latency + application latency". When $c$ has no NAS procedure, $L_c$ equals application latency. 

\begin{figure*}
  \centering
  \begin{tabular}{cc}
  \multirow{2}{*}[5cm]{\subfloat[][Network topology \cite{topology_paper}]{\label{fig:a}\includegraphics[width=.55\textwidth, scale=1]{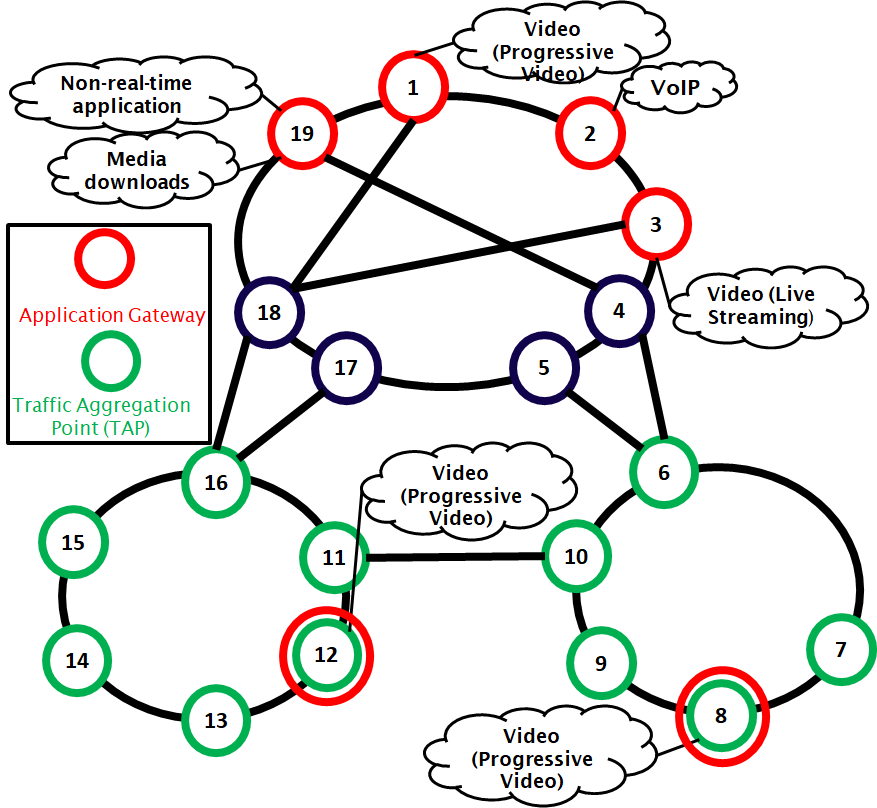}}} &
  	\begin{tabular}{c}
    	\subfloat[][Bandwidth vs. number of vEPC replicas]{\label{fig:b}\includegraphics[width=.3\textwidth, scale=1]{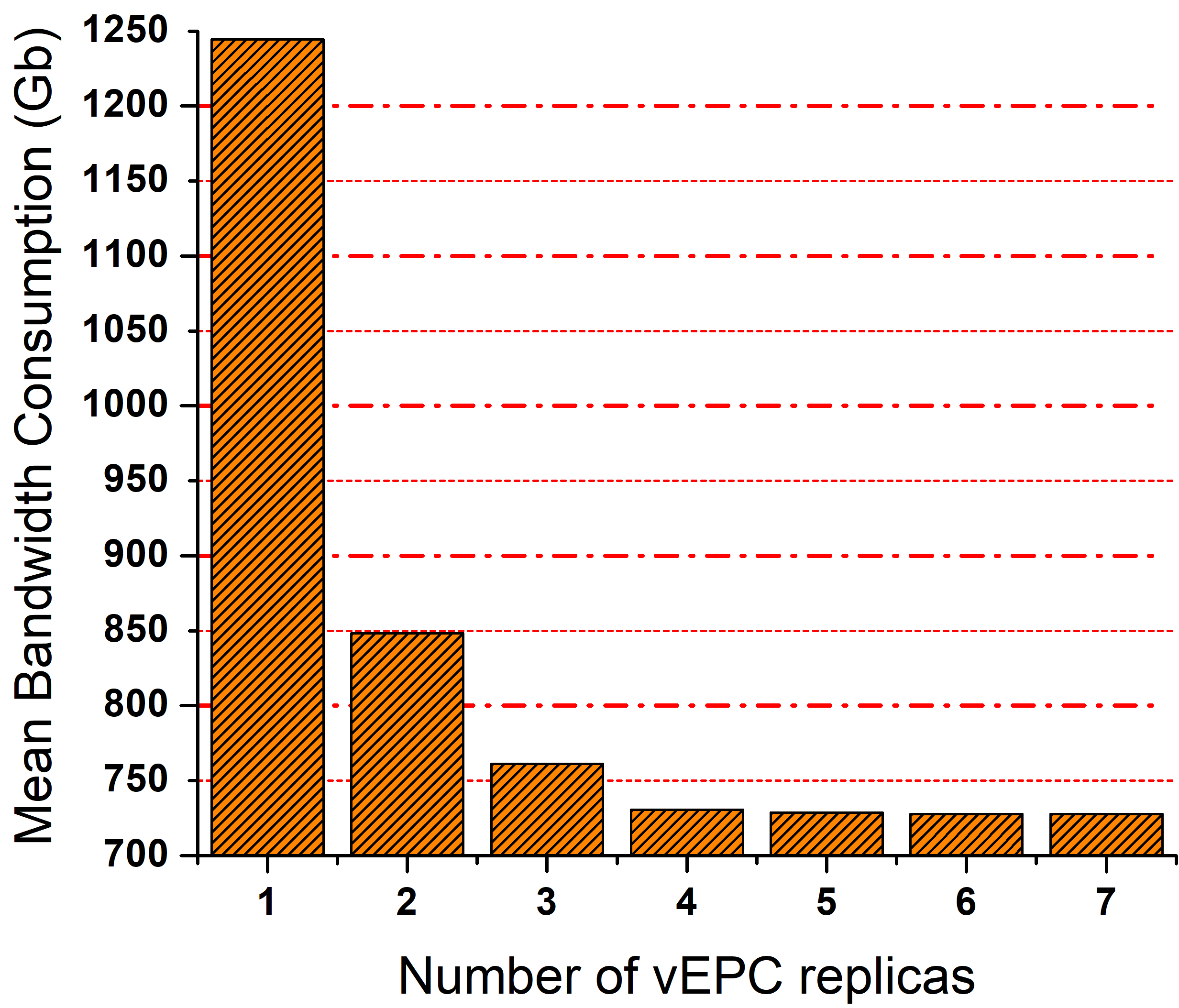}} \\
    	\subfloat[][Bandwidth vs. number of VNF replicas]{\label{fig:c}\includegraphics[width=.35\textwidth, scale=1]{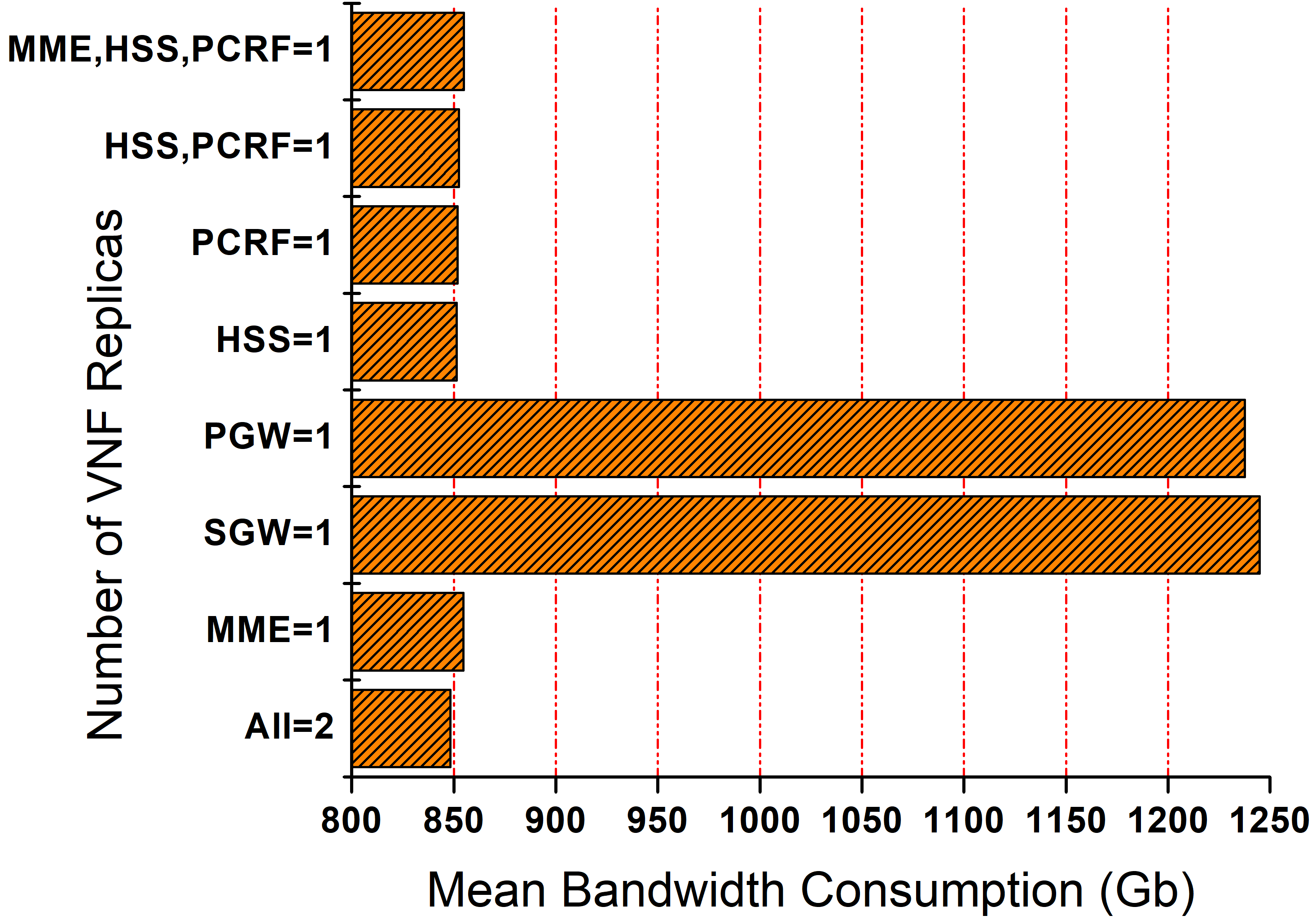}} \\
    \end{tabular} \\
  \end{tabular} 
  \caption{Simulation topology and results}
  \label{fig:results}
\end{figure*}

The Integer Linear Program (ILP) formulation of the vEPC placement problem is detailed in the following.

\textbf{Objective:} Minimize bandwidth consumed:
\begin{equation}
\sum\limits_{\ServiceChainIdx \in \ServiceChainSet}  \> \sum\limits_{\ell \in L} \> \sum\limits_{i =1}^{n_{\ServiceChainIdx} - 1} D^{\ServiceChainIdx} \beta_{i}^{c}  y_{i \ell}^{c}
\end{equation}

\noindent
Eq.(1) gives total network-resource consumed to route SCs.
\begin{alignat}{2}
& x_{v i}^{c j} = 1 
&& \nonumber  \ServiceChainIdx \in \ServiceChainSet, v \in \mathcal{SD}_c, j=\theta_i(\ServiceChainIdx)\\
& &&  i \in \{1, 2, \dots, n_{\ServiceChainIdx}\}: f_i \in \{f_s^c,f_d^c\}    \label{eq:each_function_1}\\
& x_{v i}^{c j}= 0
&& \nonumber  \ServiceChainIdx \in \ServiceChainSet, v \in \mathcal{SD}_c: v \notin V^{\NFV}, j=\theta_i(\ServiceChainIdx),\\
& &&   i \in \{1, 2, \dots, n_{\ServiceChainIdx}\}: f_i \notin \{f_s^c,f_d^c\}    \label{eq:each_function_2} \\
& \sum\limits_{v \in V^{\NFV}} x_{v i}^{c j} = 1 \quad
&& \nonumber \ServiceChainIdx \in \ServiceChainSet, j=\theta_i(\ServiceChainIdx),\\
& &&  i \in \{1, 2, \dots, n_{\ServiceChainIdx}\}: f_i \notin \{f_s^c,f_d^c\} \label{eq:each_function_3}
\end{alignat} 
Eq. \eqref{eq:each_function_1} enforces VNFs $(f_s^c,f_d^c)$ be placed on $(v_s,v_d)$ respectively while Eq. \eqref{eq:each_function_2} avoids placing vEPC VNFs on $(v_s,v_d)$ if they are not NFV nodes. Eq. \eqref{eq:each_function_3} ensures that all vEPC VNFs in $c$ are placed at exactly one node. 

The next constraints are flow conservation ones and are responsible for explicit service chaining.
\begin{alignat}{2}
&    \sum\limits_{\ell \in \omega^+(v)} y_{i \ell}^c 
    - \sum\limits_{\ell \in \omega^-(v)}  y_{i \ell}^c
      = x_{v i}^{c j} - x_{v, i+1}^{c j}  && \nonumber \\
& && \hspace*{-4cm} \ServiceChainIdx \in \ServiceChainSet ,v \in \{V^{\NFVI} \cup \mathcal{SD}_c\},  j=\theta_i(\ServiceChainIdx), \nonumber \\
& && \hspace*{-4cm}  i = 1, 2, \dots, n_{\ServiceChainIdx} - 1 \label{eq2:flowNFV} \\
& \sum\limits_{\ell \in \omega^+(v)} y_{i \ell}^c - \sum\limits_{\ell \in \omega^-(v)} y_{i \ell}^c = 0 && \nonumber \\
& && \hspace*{-4cm} \ServiceChainIdx \in \ServiceChainSet ,v \in V \setminus \{V^{\NFVI} \cup \mathcal{SD}_c\},  \nonumber \\
& && \hspace*{-4cm} i = 1, 2, \dots, n_{\ServiceChainIdx} - 1 \label{eq2:flow}
\end{alignat} 

\begin{alignat}{2}
&  M x_{vf} \geq \sum\limits_{\ServiceChainIdx \in \ServiceChainSet: f \in \ServiceChainIdx} \> \sum\limits_{\substack{ i \in \{1,2,\dots, n_c\}: \\ f_i = f, j= \theta_i(\ServiceChainIdx)}}  x_{v i}^{c j}  \geq x_{vf} 
 && \nonumber \\
& && \hspace*{-2cm} v \in V^{\NFV}, f \in F \label{eq:f_on_v} \\
& \sum\limits_{v \in V^{\NFV}} x_{v f} \leq R_f
&&  \hspace*{-2cm}  f \in F \label{eq2:each_function}
\end{alignat} 

Eqs. \eqref{eq:f_on_v} and \eqref{eq2:each_function} keep track of VNF replicas. 

\begin{alignat}{2}
& \sum\limits_{\SFCidx \in \SFCset} D^c 
    \>  \sum\limits_{i=1, j=\theta_i(\ServiceChainIdx)}^{n_c} \beta_{i-1}^{c} x_{v i}^{cj} \tfic1 \ncoref  \leq \ncore 
&&  \nonumber \\
& && \hspace*{-2cm}v \in V^{\NFV} \label{eq:capa_cores}\\
&   \sum\limits_{\ServiceChainIdx \in \ServiceChainSet} D^{\ServiceChainIdx} \> \sum\limits_{i =1}^{n_{\ServiceChainIdx} - 1} \beta_{i}^{c}  y_{i \ell}^{c} \>  \leq \textsc{cap}_{\ell}  
&& \hspace*{-2cm} \ell \in L \label{eq:capacity} \\
& \sum\limits_{\ell \in L} \> \sum\limits_{i =1}^{n_{\ServiceChainIdx} - 1} {\Delta}_{l}^{propag} y_{i \ell}^{c} + 
 \sum\limits_{i =1}^{n_{\ServiceChainIdx}} \beta_{i-1}^{c} D^c {\Delta}_{f}^{proc} \tfic1 \nonumber \\
 & \hspace*{3.cm} \leq L_c 
 &&  \hspace*{-2cm}  \ServiceChainIdx \in \ServiceChainSet.  \label{eq:latency} 
\end{alignat} 
Eq. \eqref{eq:capa_cores} bounds total CPU resource consumption by VNFs at a node. Eq. \eqref{eq:capacity} constrains total bandwidth consumed in routing SCs at a link. Eq. \eqref{eq:latency} enforces that the total latency incurred in propagation across links and processing by VNFs for each SC does not exceed latency bounds. 


\begin{table}
 \centering
 \begin{tabular}{|c|c|} 
 \hline
 Application & Traffic \% \\ [0.5ex] 
 \hline
 Progressive video (buffered streaming) & 71.19\% \\ 
 Video conferencing & 4.56\% \\
 VoIP & 1.50\% \\ 
 Media downloads & 13.3\% \\
 Non-real-time application (web,email) & 9.45\%\\ 
 \hline 
 \end{tabular}
 \caption {Application traffic \cite{nas_flows}}
 \label{table:app_traffic}
\end{table}

\section{Illustrative Numerical Examples}
\label{num_examples}

\begin{table}
 \centering
 \begin{tabular}{|c|c|c|c|} 
 \hline
 NAS procedure & Flows & Bearer & Latency (ms)\\ [0.5ex] 
 \hline
 Attach & 10 & Default & 500\\ 
 Dedicated bearer request & 45 & Dedicated & 250\\
 X2-based handover & 5 & Default & 500\\ 
 S1-based handover & 10 & Default & 500\\ 
 \hline 
 \end{tabular}
 \caption {Traffic flows with NAS procedure requirements \cite{nas_flows}}
 \label{table:flow_count_ctrl_lat}
\end{table}

We run our simulations on the 19 node topology shown in Fig. \ref{fig:results}\subref{fig:a}. The application gateways shown in Fig. \ref{fig:results}\subref{fig:a} are selected based on Table \ref{table:app_traffic} and are shown in red, green nodes are TAPS and black nodes are switches. We add Multi-Access Edge Computing (MEC) sites at node 12 and 8 for progressive video. This decision is taken as progressive video has highest traffic load, which MEC can reduce. Table \ref{table:flow_count_ctrl_lat} shows the number of application traffic flows for NAS procedures ( aggregated from 1000 to 5000 UEs).  These traffic flows are associated with a upload or download DSC. We split the application traffic in Table \ref{table:app_traffic} over the flows shown in Table \ref{table:flow_count_ctrl_lat} and 50 flows with no NAS procedure requirements (i.e., only upload/download). We consider the upload to download traffic ratio as 1:4 \cite{itu_ratio}. Latency requirements for applications are taken from \cite{app_latency} and for NAS procedures depends on bearer type, as seen in Table \ref{table:flow_count_ctrl_lat}.


Total traffic is 224 Gb \cite{nas_flows}. All nodes are allowed to host VNFs. Each node is allocated 2400 CPUs. Link capacity is set to 60 Gbps. Control plane traffic is 5\% of data plane traffic i.e. $B^c_i=0.05$ for control plane signaling. For data path $B^c_i=1.0$ since all traffic is data traffic. All links are considered optical fibers of length 50km. Processing latency is 132 $\mu$s per Gbps \cite{basta_processing} for each VNF. All VNFs require 2 CPUs per Gbps of throughput \cite{brocade_vepc}.

We run 10 iterations of our optimization model and plot mean results. Fig. \ref{fig:results}\subref{fig:b} shows reduction in bandwidth consumption as the number of allowed vEPC replicas (Eq. \eqref{eq2:each_function}) are increased. We find the largest reduction occurs from 1 to 2 vEPC replicas. This happens because vEPC gets distributed across aggregation rings which reduces route length to the core, thereby resulting in reduction of bandwidth consumption. It should be noted that MEC is a significant factor in bandwidth usage reduction since applications become available at the edge. vEPC without MEC would always require routing to the core and reduction in bandwidth consumption would not be significant. We find that as number of replicas are increased beyond 4, reduction is not significant as we reach almost optimal bandwidth consumption by 4 replicas. Having 2 vEPC replicas reasons to be the best tradeoff between bandwidth reduction and the overhead of deploying more replicas and making more nodes NFV compatible. 

Figure \ref{fig:results}\subref{fig:c} shows bandwidth consumption for different number of VNF replicas. Here, all VNFs have 2 replicas unless specified on the Y-axis. This tells us that having 1 replica of MME, HSS and PCRF has a negligible effect on bandwidth consumption in comparison to having 2 replicas of each (All=2). Hence, we only need to deploy 2 replicas of PGW and SGW to achieve the same bandwidth reduction as 2 vEPC replicas in Fig. \ref{fig:results}\subref{fig:b}. This demonstrates that we do not need to replicate all VNFs to achieve bandwidth reduction. 

\section{Conclusion}
\label{concl}
We introduce the problem of vEPC placement while accounting for VNF interactions in control plane and data plane and application latency. We develop an Integer Linear Program (ILP) for placement of EPC VNFs and route traffic along service chains. We demonstrate that there is reduction in network-resource consumption with increase in number of VNF replicas. Further, we show that not all EPC VNFs need to be replicated and that SGW and PGW replication gives almost the same network resource consumption as replicating all vEPC VNFs.


\section*{Acknowledgment}
This work was supported by NSF Grant No. CNS-1217978.



%
\bibliographystyle{IEEEtran}
\bibliography{gupta_nfv_epc}

\begin{thebibliography}{10}
\providecommand{\url}[1]{#1}
\csname url@samestyle\endcsname
\providecommand{\newblock}{\relax}
\providecommand{\bibinfo}[2]{#2}
\providecommand{\BIBentrySTDinterwordspacing}{\spaceskip=0pt\relax}
\providecommand{\BIBentryALTinterwordstretchfactor}{4}
\providecommand{\BIBentryALTinterwordspacing}{\spaceskip=\fontdimen2\font plus
\BIBentryALTinterwordstretchfactor\fontdimen3\font minus
  \fontdimen4\font\relax}
\providecommand{\BIBforeignlanguage}[2]{{%
\expandafter\ifx\csname l@#1\endcsname\relax
\typeout{** WARNING: IEEEtran.bst: No hyphenation pattern has been}%
\typeout{** loaded for the language `#1'. Using the pattern for}%
\typeout{** the default language instead.}%
\else
\language=\csname l@#1\endcsname
\fi
#2}}
\providecommand{\BIBdecl}{\relax}
\BIBdecl

\bibitem{nas_procs}
\BIBentryALTinterwordspacing
3GPP, ``{3GPP TS 24.301: 3GPP Non-Access-Stratum (NAS) protocol for Evolved
  Packet System (EPS).}'' [Online]. Available:
  \url{http://www.3gpp.org/DynaReport/24301.htm.}
\BIBentrySTDinterwordspacing

\bibitem{rajan_nas_paper}
{A. S. Rajan et al.}, ``Understanding the bottlenecks in virtualizing cellular
  core network functions,'' in \emph{IEEE Workshop on LANMAN}, 2015.

\bibitem{ietf_sc}
IETF, ``Network service chaining problem statement,''
  \url{https://tools.ietf.org/html/draft-quinn-nsc-problem-statement-00}, 2013.

\bibitem{gupta_gc17}
A.~Gupta, B.~Jaumard, M.~Tornatore, and B.~Mukherjee, ``{Service Chain (SC)
  Mapping with Multiple SC Instances in a Wide Area Network},'' in \emph{IEEE
  GLOBECOM 2017}, Dec 2017, pp. 1--6.

\bibitem{gupta_jsac18}
{A. Gupta, B. Jaumard, M. Tornatore and B. Mukherjee}, ``{A Scalable Approach
  for Service Chain (SC) Mapping with Multiple SC Instances in a Wide-Area
  Network},'' \emph{IEEE Journal on Selected Areas in Communications}, 2018.

\bibitem{etsi_mec}
\BIBentryALTinterwordspacing
ETSI, ``{Mobile-Edge Computing - Introductory Technial White Paper}.''
  [Online]. Available:
  \url{https://portal.etsi.org/portals/0/tbpages/mec/docs/mobile-edge_computing_-_introductory_technical_white_paper_v1%2018-09-14.pdf}
\BIBentrySTDinterwordspacing

\bibitem{baumgartner_vepc}
A.~Baumgartner, V.~S. Reddy, and T.~Bauschert, ``Mobile core network
  virtualization: A model for combined virtual core network function placement
  and topology optimization,'' in \emph{IEEE NetSoft 2015}.

\bibitem{baras_vepc}
D.~Dietrich, C.~Papagianni, P.~Papadimitriou, and J.~S. Baras, ``Network
  function placement on virtualized cellular cores,'' in \emph{COMSNETS 2017}.

\bibitem{topology_paper}
F.~Z. Yousaf, J.~Lessmann, P.~Loureiro, and S.~Schmid, ``{SoftEPC - Dynamic
  instantiation of mobile core network entities for efficient resource
  utilization},'' in \emph{IEEE ICC 2013}.

\bibitem{taleb_relocation}
T.~Taleb and A.~Ksentini, ``{Gateway Relocation Avoidance-aware Network
  Function Placement in Carrier Cloud},'' in \emph{Proceedings of the 16th ACM
  MSWiM '13}.

\bibitem{nas_flows}
B.~H. et~al., ``High-performance evolved packet core signaling and bearer
  processing on general-purpose processors,'' \emph{IEEE Network}, vol.~29,
  no.~3, pp. 6--14, 2015.

\bibitem{itu_ratio}
\BIBentryALTinterwordspacing
ITU, ``{Report ITU-R M.2370-0: IMT traffic estimates for the years 2020 to
  2030.}'' [Online]. Available:
  \url{https://www.itu.int/dms_pub/itu-r/opb/rep/R-REP-M.2370-2015-PDF-E.pdf}
\BIBentrySTDinterwordspacing

\bibitem{app_latency}
3GPP, ``{3GPP TR23.401 V8.1.0: General Packet Radio Service (GPRS) enhancements
  for Evolved Universal Terrestrial Radio Access Network (E-UTRAN) access}.''

\bibitem{basta_processing}
A.~Basta, W.~Kellerer, M.~Hoffmann, H.~J. Morper, and K.~Hoffmann, ``{Applying
  NFV and SDN to LTE mobile core gateways, the functions placement problem},''
  in \emph{ACM Proceedings of the 4th workshop on All things cellular:
  operations, applications, \& challenges 2014}, pp. 33--38.

\bibitem{brocade_vepc}
\BIBentryALTinterwordspacing
Brocade, ``{Brocade vEPC}.'' [Online]. Available:
  \url{https://www.brocade.com/content/dam/common/documents/content-types/datasheet/brocade-vepc-ds.pdf}
\BIBentrySTDinterwordspacing

\end{thebibliography}

\end{document}